\documentstyle[aps,preprint]{revtex}
\begin{document}
\draft

\title{Peculiarities in Low Temperature Properties of 
Doped Manganites A$_{1-x}$B$_x$MnO$_3$.}
\author{M.\ O.\ Dzero$^{1}$, L.\ P.\ Gor'kov$^{1,2}$ and 
V.\ Z.\ Kresin$^{3}$}
\address{$^1$National High Magnetic Field Laboratory, Florida State
University, Tallahassee, FL 32310}
\address{$^2$L.D. Landau Institute for Theoretical Physics, Russian Academy
of Sciences, 117334 Moscow, Russia}
\address{$^3$Lawrence Berkeley Laboratory, University of California, Berkeley, 
CA 94720}
\date{\today}
\maketitle

\begin{abstract}
The phase diagram and low temperature properties of the doped
manganites A$_{1-x}$B$_{x}$MnO$_3$  are discussed for the
concentrations $x\le 0.4$. The transition from insulating
antiferromagnetic to metallic ferromagnetic state at $x_{cr}
\simeq 0.16$ is treated by means of percolation theory.
The unifying description of insulating and metallic states
is presented. The undoped manganite is a band insulator 
consisting of ferromagnetic layers, which are coupled 
antiferromagnetically along the $c$ direction with
a low N$\acute e$el temperature. The metallic phase
can be described by the two-band Fermi liquid picture.
The behavior of conductivity, spin wave excitations, etc.
is analyzed and the comparison with experimental data is
carried out.
\end{abstract}
\section{Introduction}
Interest in utilizing the ``colossal magnetoresistance'' (CMR)
effect has resulted in enormous literature,
which spans properties of manganites at numerous compositions,
ceramic and crystals, crystalline films on different 
substrates, stoichiometric and not (for extensive
review see \cite{vmolnar,ramirez}).
Nevertheless, until recently, it was somewhat difficult to
establish systematic trends which govern the physical properties
of these materials (especially their low temperature behavior), 
to access peculiarities pertinent to the ground state and
its dependence on composition.

As for the overall behavior, especially for the high temperature
properties, which are responsible for the CMR phenomena itself,
the consensus is that the Zener's double exchange (DE) mechanism
\cite{Zener1,Zener2,Anderson} together with the Jahn-Teller (JT) effect for
Mn$^{3+}$ as suggested in \cite{Millis1,Millis2} 
produce the basis for
their understanding. High temperature lattice deformations
localize carriers by a formation of the JT or local polarons
\cite{Millis1,Roeder}. With temperature decrease, DE mechanism delocalizes 
carriers and result in simultaneous onset of metallic conductivity
and half-metallic ferromagnetic state. The physical picture, 
is by all means correct, but the method, 
introduced in \cite{Millis1}, 
is not sufficient for understanding delicate 
qualitative features, especially at the low temperatures.

As an important issue, we have to mention, is the evolution
of ground state for the doped manganites from the insulating
antiferromagnetic state of LaMnO$_3$ to the pronounced 
metallic ferromagnetic one in La$_{1-x}$Sr$_x$MnO$_3$ at $x\sim 0.3$.

It has been demonstrated first in our short paper \cite{LevandKresin} 
that both
the half-metallic state at finite concentrations and the insulating
state of the parent LaMnO$_3$ can be brought together by
making use of a simple band model. Parent LaMnO$_3$ is considered
in that scheme as a band insulator. As for the doping process,
it has been shown in \cite{LevandKresin}, that it leads to the percolative
scenario, producing the well defined threshold concentration
$x_{cr}\sim 0.16-0.17$ for the onset of low-temperature metallic
conductivity.

In the presentation below we describe in more details 
and further elaborate the theory, proposed by Gor'kov and
Kresin, keeping in mind both some future applications 
and analysis of available experimental data at $x\leq 0.4$ 
is also provided.
It is shown that the two bands theory agrees reasonably well
with the results of recent experiments, demonstrating
applicability of the Fermi liquid approach to the metallic
manganites. The scheme serves as a theory unifying rather 
rich low temperature properties of manganites and is capable
of some predictions regarding the evolution of physical parameters,
the shape of the Fermi surface etc. with concentration.
The experimental data show also the consistency with our results
in the metallic concentration range and trace remaining 
percolation features. 

The paper is written as follows. In Sec. II  the main features 
are identified and accounted in Hamiltonian. Sec. III contains
the energy spectrum and its dependence on a spin ordering or
JT - deformations. Properties of a stoichiometric compound LaMnO$_3$,
treated as a band insulator, are considered in Sec. IV.
Growth of the percolative regime for the phase separation
and the critical composition for onset of metallic conduction are
discussed in Sec. V. Sec. VI contains a summary of major metallic 
properties calculated in the framework of the two band model,
while in Sec. VII a comparison with available experimental data
is discussed. Sec. VIII contains brief summary and conclusions. 

\section{Model and Hamiltonian}
We just start with, a somewhat oversimplified
(the so-called ``pseudocubic'') crystalline structure
for manganites. As it is well-known ~\cite{vmolnar,ramirez},
in that approximation the unit cell, say, for LaMnO$_3$,
may be taken as a cube, with the lattice constant
$a_0 \simeq 3.9 \AA$. Rare-earth or alkaline ions are placed
at the center , while the manganese ions occupy 
the corner sites. Mn$^{3+}$-- sites are caged into the
oxygen octahedra, which share the O$^{2-}$ ions along
the Mn-O-Mn bond (another view would be that each
 La$^{3+}$-- ion is cooped up in the midst of twelve 
 O$^{2-}$ ions). The ideal structure is then modified for
real materials, AMnO$_3$, due to a mismatch in the ionic radii.
The latter is commonly characterized by the tolerance factor,
$t$ (see e.g. in ~\cite{vmolnar}):
\begin{equation}
t = \frac{1}{\sqrt{2}}\cdot \frac{R_A + R_O}{R_{Mn} + R_O}.
\label{tol}
\end{equation}
The effect of $t\not= 1$ is that the oxygen octahedra become
periodically tilted, and the unit cell may then be comprised
of a few ``pseudocubic'' cells. It is shown below that 
deviations from the ``pseudocubic'' structure, i.e. deviations in
the angle, $\alpha$, of the Mn-O-Mn bond from $180^{\circ}$
are not of much importance for the ``average'' electronic structure.
However, we will see later that {\it local} fluctuations in the
tolerance factor (\ref{tol}) may play rather significant
role for the conducting properties of the ``doped'' manganites
A$_{1-x}$B$_{x}$MnO$_3$.

Let us outline our base model which is
then applied for the interpretation of manganites properties.
Since the manganese $d$- shell has reasonably small orbitals,
the model exploits the tight binding approximation.

As it is well known, the Mn$^{3+}$ ion has four $d$-- electrons.
The $d$-- shell in the cubic environment is split into the
triple ($t_{2g}$) and double ($e_{2g}$) degenerate terms.
The $t_{2g}$-- level is fully occupied by three electrons considered
as a local spin S = $3/2$, in accordance with the Hund's rule.
The $e_{2g}$-- term in manganites may be empty (Mn$^{4+}$) or
single occupied (Mn$^{3+}$). If there is one electron on the
$e_{2g}$-- level, the direction of its spin is governed by the same
intra-atomic Hund's interaction:
\begin{equation}
\hat H_H = - J_H {~\bbox{\sigma}}\cdot {\bf S}_i,
\label{dva}
\end{equation}
($\bbox{\sigma}$-- the Pauli matrix for the $e_{2g}$-- electron).
The Hund's coupling $J_H > 0$ is rather large 
($J_H S \sim 1 \div 1.5 eV$). Therefore spins of $e_{2g}$ and
$t_{2g}$ electrons are ferromagnetically aligned locally.

The two major assumptions are: 1. -- the splitting between the 
$t_{2g} - $ and $e_{2g} -$ levels is large enough, so that
only $e_{2g}$ electrons may participate in conductivity;
2. -- the $e_{2g} -$ levels are well above the top of the filled
up oxygen band (on a scale of a few eV).
The band-structure calculations do not always support the last assertion
 (see e.g. ~\cite{pickett}). On the other hand, extensive optical
studies agree well with the second assumption, see e.g. \cite{optic}.

Hoping process of one electron from a site $i$ to its nearest neighbor
$i+\delta$ in the tight-binding approximation has to be modified
to account for the double degeneracy of the $e_{2g}$ - level 
on each site. In the hopping Hamiltonian, hence:
\begin{equation}
\hat H_t = \sum\limits_{i, \delta}^{} \hat t_{i, i+\delta},
\label{hop}
\end{equation}
$\hat t_{i, i+\delta}$ becomes a two-by-two matrix. Its explicit
form  depends
on choice of the basis for the $e_{2g}$ - representation.

The $e_{2g}$ - degeneracy of the Mn$^{3+}$ site would locally
result in the Jahn - Teller (JT) instability with respect
to spontaneous lattice distortions. Two active modes for deformation
of the surrounding oxygen octahedron are usually denoted in
literature as $Q_2$ and $Q_3$ (see \cite{Kanamori} and review
\cite{kugel}), and form
the two-dimensional basis of the octahedron deformations. In
the invariant form the JT part of the electron-lattice interaction
may be written as:
\begin{equation}
H_{JT} = g ~\bbox{\hat{\tau_i}}\cdot {\bf Q}_i,
\label{JT}
\end{equation}
where $\bbox{\hat{\tau}}$ is called the ``pseudospin'' matrix
(see ~\cite{kugel}).

In what follows we use the normalized complex basis 
\cite{LevandKresin,Levufn} for the two-dimensional
representation $e_{2g}$:
\begin{equation}
\psi_1\propto z^2+\epsilon x^2+\epsilon^2y^2; ~ \psi_2\equiv\psi_1^{\ast},
\label{wfunct}
\end{equation}
where $\epsilon =\exp (2\pi i/3)$. 
Another basis, which is often used in literature,
for example ~\cite{kugel,khomskii},
is given by the real functions:
\begin{eqnarray}
\varphi_1 \propto d_{z^2} \equiv 3 z^2 - (x^2 + y^2);
 ~\varphi_2 \propto d_{x^2 - y^2} \equiv x^2 - y^2.
\eqnum{5'}
\label{eq:5p}
\end{eqnarray}
Connection between these two is:
\begin{eqnarray}
\psi_1 = (\varphi_1 + i \varphi_2)/{\sqrt{2}}; 
 ~\psi_2\equiv\psi_1^{\ast}.
\eqnum{5''}
\label{eq:5pp}
\end{eqnarray}
The given basis (\ref{wfunct}) makes it possible
to present (\ref{JT}) in the convenient form:
\begin{eqnarray} 
-\frac{g}{2}Q_0\left( \begin{array}{cc} 0 & \exp (i\theta) \\ \exp
(-i\theta) & 0 \end{array}\right)
\end{eqnarray}
where in the standard notations ~\cite{Kanamori,kugel,khomskii}:
\begin{equation} 
Q_2 = Q_0 \sin\theta ~;~ Q_3 = Q_0 \cos\theta
\label{qbas}
\end{equation}
with $Q_0$ being the magnitude of the JT - distortions.
The angle, $\theta$, specifies the shape of the distorted octahedron.
Thus, the angles $\theta = 0, \pm 2\pi/3$ correspond to elongation
of the octahedron along the z, x and y - axes, respectively.

The JT - term (\ref{JT}) is linear in $Q$, while the elastic energy
is quadratic in $Q$. Therefore if one electron is placed on
the JT - level, the total site energy always decreases with 
non-zero  lattice deformations. 
Strictly speaking, deformations, $Q_i$, on the two
adjacent manganese sites are not independent because the two sites
share one oxygen along the Mn-O-Mn bond;
in what follows only cooperative JT distortions will be considered.

According to review \cite{Kanamori} (see also \cite{kugel})
we are dealing with two
normal modes $Q_2$ and $Q_3$, where:
\begin{eqnarray}
\begin{array} {c} Q_2 = \frac{1}{\sqrt{2}}(x_1 - x_4 + y_5 - y_2),\\
Q_3 = \frac{1}{\sqrt{6}}(2 z_2 - 2 z_6 - x_1 - x_4 - y_2 + y_5).
\end{array}
\label{eq:8}
\end{eqnarray}

Onset of the cooperative JT distortions means the onset of a non-cubic
structural order (``ferrodistorsive'' deformations, if the wave
vector of the structure, ${\bf q}=0$, ``antiferrodistorsive'' ones,
if ${\bf q}\neq 0$ ~\cite{Kanamori,Kaplan}). To account for fluctuations or
changes in the phonon spectrum, the displacements in (\ref{eq:8})
must be expressed in terms of the normal modes for the crystal
vibrations.

With the above remark regarding the JT-distortion, one may now
bring all the contributions (\ref{dva},\ref{hop},\ref{JT}) together:
\begin{equation}
H=\sum\limits_{i}^{}
\left (\sum\limits_{\delta}^{}\hat{t}_{i, i+\delta} - J_H
{\bbox{\sigma}}\cdot{\bf S}_i + g{\bbox{\hat{\tau}}}_i \cdot 
{\bf Q}_i + J_{el} {\bf Q}_{i}^{2} \right )
\label{eq:9}
\end{equation}
(the last term added into (\ref{eq:9}) is responsible for 
the ``elastic'' energy of the JT mode). The Hamiltonian of
exactly the form (\ref{eq:9}) has been considered by Millis et. al.
~\cite{Millis1,Millis2}. As it was mentioned above, the important
physical observation made in ~\cite{Millis1,Millis2} was that with the
temperature increase disorder produced by thermally excited 
JT degrees of freedom tends to localize the charge carriers.
Though the idea by all means is qualitatively correct, the approach, used
in ~\cite{Millis1}, is a selfconsistent interpolative
scheme which is not appropriate for understanding low
temperature properties of manganites, i. e. the symmetry of the
ground state and its variation with composition. The latter 
issues are addressed below mainly in terms of the band approach.
\section{Band Spectrum}
Let us consider first the band spectrum of the Hamiltonian
(\ref{hop},\ref{JT}) (the Hund's term and JT term being temporary omitted).
The matrix $\hat{t}$ in (\ref{hop}) on the basis (\ref{wfunct})
has the form:
\begin{eqnarray}
\left( \begin{array}{cc} {\Sigma}_{11} & {\Sigma}_{12} \\
{\Sigma}_{21} & {\Sigma}_{22} \end{array}\right),
\label{eq:10}
\end{eqnarray}
where 
\begin{eqnarray}
\begin{array}{c} \Sigma_{11} = \Sigma_{22} = \Sigma_0 = (A+B)
[\cos(k_x a) + \cos(k_y a) + \cos(k_z a) ], \\
\Sigma_{12} = \Sigma_{21}^{*} = (A-B) 
[\cos(k_z a) + \epsilon\cos(k_x a) + {\epsilon}^2 \cos(k_y a)].
\end{array}
\label{eq:11}
\end{eqnarray}
A and B in (\ref{eq:11}) are two overlap integrals:
\begin{eqnarray}
\begin{array}{c} A \propto 
\overline{\varphi_1(z;x,y)\varphi_1(z+a;x,y)}, \\ 
B \propto 
\overline{\varphi_2(z;x,y)\varphi_2(z+a;x,y)}, \\
\end{array}
\label{eq:12}
\end{eqnarray}
where the bar $\overline{(...)}$ means the matrix elements for
the interaction potential on the two Wannier functions, $\varphi_{1,2}$,
of the neighboring atoms (in the tight binding approximation
$A = - |A| < 0$). Simple geometric considerations for the d-shell
show that $|B| < |A| (|B| \simeq \frac{1}{16} |A|$, according 
to \cite{Godunov,Anderson1}). The cubic spectrum consists of the
two branches:
\begin{equation}
{\varepsilon}_{1,2}({\bf p}) = - (|A| + |B|)(c_x+c_y+c_z) \pm
(|A| - |B|)\sqrt{c_x^2 + c_y^2 + c_z^2 - 
c_x c_y - c_y c_z - c_z c_x}
\label{eq:13}
\end{equation}
(we introduced the notations $c_i = \cos(k_i a), i=x,y,z$).

If a homogeneous ``ferroelastic'' JT-distortion is imposed on the
lattice, the spectrum may be obtained by adding $-\frac{g Q_0}{2}
\exp{(i \theta)}$ term to $\Sigma_{12}$ in (\ref{eq:10}). For some
future applications we provide here the spectrum for the limiting
case $|g Q_0| \gg |A|$:
\begin{eqnarray}
\widetilde{\varepsilon}_{1,2}({\bf p})&=&\pm \mid\frac{gQ_0}{2}\mid
+ (A+B) (c_x + c_y + c_z) \pm \nonumber \\
&& (A - B) \left \{(c_z - \frac{1}{2}(c_x + c_y)) \cos{(\theta)} + 
\frac{\sqrt{3}}{2} (c_y - c_x) \sin{(\theta)} \right \}.
\label{eq:14}
\end{eqnarray}
This expression is interesting: at the choice $\sin{(\theta)} = 0$
the low-lying electronic branch would become strongly anisotropic
($|B| \ll |A|$):
\begin{equation}
\widetilde{\varepsilon_{2}} ({\bf p}) = -\mid\frac{gQ_0}{2}\mid - 
2 B c_z + \frac{3}{2} A (c_x + c_y).
\eqnum{14'}
\label{eq:14p}
\end{equation} 

Returning to the full electronic Hamiltonian, let us recall that
the Hund's term (\ref{dva}) is the largest one:
we assume below $J_H \gg |A|, g Q$. Therefore in all cases 
the electronic spectrum is shifted up or down by the energy 
$\pm J_H S$, depending on the $e_{2g}$ and $t_{2g}$ {\it mutual} 
spin direction. At the ferromagnetic alignment of the all local
and itinerant spins {\it each of} the two branches 
of Eq. (\ref{eq:13})
merely splits into two by adding the $\pm J_H <S>$ energy term.
For an antiferromagnetic (AF) ground state one needs to calculate
the energy spectrum again. It is done below for the important 
case of two ``canted'' antiferromagnetic sublattices when 
a ferromagnetic spin component ${\bf M}_i$ coexists with the 
staggered magnetization, ${\bf S}_i = (-1)^i {\bf S}_i$.
We assume that antiferromagnetic ordering with the structure
vector ${\bf q}$, runs along the $c$-axis,
${\bf q}=\frac{\pi}{a}(0,0,1)$. Such an AF - structure is often
dubbed as the A-phase.

We will perform our calculations treating local ${\bf M}_i$ 
and ${\bf S}_i$ as the classical vectors:
$({\bf M}_i + {\bf S}_i)^2 = {\bf S}^2; \mid{\bf S}\mid=3/2$.
The $e_{2g}$- electron wave function in the momentum representation 
acquires the form:
\begin{eqnarray}
\Psi_{\bf p} \equiv \left \{ \begin{array}{c} 
\alpha_{{\bf p} \sigma} \psi_1 \\ 
\beta_{{\bf p} \sigma} \psi_2 \end{array} \right \} => 
\left \{ \begin{array}{c}
\alpha_{{\bf p} \sigma} \\
\beta_{{\bf p} \sigma} \end{array} \right \},
\label{eq:15}
\end{eqnarray}
where now $\alpha_{{\bf p} \sigma},\beta_{{\bf p} \sigma}$ have
a nontrivial spin-index dependence. 
Let ${\bf S}({\bf q})$, the staggered magnetization component,
be directed along the $z$-direction, while ${\bf M}$ is parallel
to the $x$ - axis (in the spin space). 
In the calculations below we suppose
${\bf q}$ is along $c$-axis. 
The Brillouin zone
is reduced along the $c$ - direction, in accordance with
a choice of {\bf q} - vector. 
The energy spectrum is determined by
the following equations:
\begin{eqnarray}
\hat{E} \psi_{{\bf p} \sigma}&=& (\hat{t}({\bf p}) - J_H \hat{\sigma}_x M )
\psi_{{\bf p} \sigma} - 
J_H \hat{\sigma}_z S \psi_{{\bf p + q} \sigma}, \nonumber \\
\hat{E} \psi_{{\bf p + q} \sigma}&=& (\hat{t}({\bf p + Q}) - 
J_H \hat{\sigma}_x M )
\psi_{{\bf p + q} \sigma} - J_H \hat{\sigma}_z S \psi_{{\bf p} \sigma}.
\nonumber
\end{eqnarray}
Multiplying each equation by $\hat{\sigma}_z$ and eliminating 
$\psi_{{\bf p+q} \sigma}$, one arrives to the following
two-by-two matrix equation:
\begin{eqnarray}
\left [\hat{E} - \hat{t}_{\bf p+q} - J_H \hat{\sigma}_x M \right ]
\left [\hat{E} - \hat{t}_{\bf p} + J_H \hat{\sigma}_x M \right ]
\psi_{\bf p} = J_H^2 S^2 \psi_{\bf p} 
\nonumber
\end{eqnarray}
or
\begin{eqnarray}
\left [E^2 - 2 \hat{t}_{p_x, p_y} E + 2 J_H \hat{\sigma}_x M 
\hat{t}_{p_z} + \hat{t}^{2}_{p_x, p_y} - 
\hat{t}^{2}_{p_z} \right ] \psi_{\bf p} = 
J_H^2 S^2 \psi_{\bf p},
\label{eq:16}
\end{eqnarray} 
where we have separated in the $\hat{t}$ - matrix its dependence
on the momenta parallel and perpendicular to the direction
of the ${\bf q}$ - vector:
\begin{eqnarray}
\hat{t} = \hat{t}_{p_x, p_y} + \hat{t}_{p_z}
\nonumber
\end{eqnarray}
and also we have used that $\hat{t}_{p_z + q_z} \equiv - \hat{t}_{p_z}$.
Equation (\ref{eq:16}) gives $E_0 = \pm J_H S$ in the zeroth
approximation, as expected. Taking $E = -J_H S + \varepsilon$,
we obtain:
\begin{eqnarray}
\left [ \varepsilon^2 - 2 J_H S \varepsilon  + 2 \hat{t}_{p_{p_x, p_y}}
(J_H S - \varepsilon) + 2 J_H \hat{\sigma}_x M \hat{t}_{p_z} + 
\hat{t}_{p_x,p_y}^2 - \hat{t}_{p_z}^2 \right ] \psi_{\bf p} = 0.
\eqnum{16'}
\label{eq:16pr}
\end{eqnarray}
For the next approximation $(\varepsilon \sim \hat{t})$ in 
(\ref{eq:16pr}):
\begin{eqnarray}
\hat{\varepsilon}\psi_{\bf p} = \left [\hat{t}_{p_x, p_y}
+ \hat{\sigma}_x \frac{M}{S}\hat{t}_{p_z} \right ] \equiv
\left [\hat{t}_{p_x, p_y} + \hat{\sigma}_x \cos{(\theta/2)}
\hat{t}_{p_z} \right ] \psi_{\bf p},
\label{eq:17}
\end{eqnarray}
i.e. one obtains the generalized Anderson-Hasegawa 
term for the motion
along the $c$ - axis. Eq. (\ref{eq:17}) is the four-by-four matrix:
two-by-two due to the orbital degeneracy and two-by-two due to 
the spin variables. After diagonalization over the spin variables,
it provides the following tunneling matrix:
\begin{equation}
\hat{\varepsilon}_{\pm}^{(1)} = \hat{t}_{p_x, p_y} \pm 
\cos{(\theta/2)} ~\hat{t}_{p_z}.
\label{eq:18}
\end{equation}  
To obtain the new spectrum one merely has to make a 
substitution in Eq. (\ref{eq:13}): 
$c_z \rightarrow \cos{(\frac{\theta}{2})} c_z$ (the
number of energy branches is doubled because of reduction
of the Brillouin zone at doubling the periodicity along the 
$c$ - direction.
If, in addition, there is a homogeneous JT deformation, all
$\hat{t}_{p_x, p_y}$ and $\hat{t}_{p_z}$ must be changed in the
same manner as it has been done above). In accordance with 
the DE - mechanism,
any connection between adjacent ferromagnetic layer in the
$A$-phase fully disappears in the
approximation (\ref{eq:18}). To find further approximations,
one has to take into account small terms of the order of
$t^2 / J_H \ll t$. The analysis below produces the scale
for $t$ in (\ref{eq:13}): $\mid{A}\mid \sim 0.16 eV $, i. e. these
terms should be reasonably small and may be neglected 
except for some more peculiar phenomena in doped manganites.
Regarding the magnitude of the JT-component, Eqs. (\ref{JT},
\ref{qbas}), it is suggested below that main physics to large
extent depends on competition between the energy gained in
the JT-distortions and kinetic (band) energy of electrons.
Correspondingly, the JT-term is generally assumed to be of the same 
order as $\mid{A}\mid$,
$ g Q_0 \sim \mid{A}\mid$, in our model.

\section{Parent manganite as a band insulator}

At low temperatures the stoichiometric ends of materials
in series A$_{1-x}$B$_{x}$MnO$_3$ are insulators.
For the ``right'' end ($x=1$) it causes no surprise:
all manganese ions are Mn$^{4+}$, i.e. comprise of the
localized $t_{2g}$ - spins. The magnetic properties are then
described by an effective Heisenberg Hamiltonian commonly
attributed to the superexchange mechanisms along the Mn-O-Mn bonds
(the Hamiltonian itself may be constructed by the so-called
``Anderson-Goodenough-Kanamori rules'' (see in 
\cite{Anderson,Kanamori,Godunov}).  For the same
approach to be applied for the ``left'' end ($x=0$, 
LaMnO$_3$), one should similary treat the Mn$^{3+}$ 
$e_{2g}$ - shells as localized spin orbitals. 
All pertinent properties are often interpreted in terms of generalized
microscopic Hubbard model (see e.g. \cite{kugel,khomskii}).
The key feature of the Hubbard model is the assertion that
for two electrons to be placed on the same site the energy cost
is very high (the famous Hubbard $ ``U'' > 0$ due to the on-site
Coulomb repulsion!).

The Hubbard Hamiltonian approach has been challenged in \cite{LevandKresin}. 
First, there are experimental motivations for such challenge.
For instance it was shown than in ``doped'' manganites,
La$_{1-x}$Sr$_{x}$MnO$_3$, at rather low concentrations, say,
$x=0.2$, the system displays
excellent low $T$ metallic behavior \cite{Lofland1}. Meanwhile, the nominal
number of $e_{2g}$-electrons per $Mn$ site, $N = 1 - x < 1$,
was changed only by one fifth in this study. From the theoretical
side, it is of course clear that dealing with the strictly atomic 
$d$-orbitals would be a strong oversimplification. If a Mn ion
is placed into the oxygen octahedron environment, the 
$e_{2g}$-terms are formed by the whole ligand, so that the ``pure''
$d$-functions become considerably hybridized with the surrounding
oxygen states (e.g. see the discussion in \cite{pickett,Anderson1}).
Hence the electronic polarization would undoubtly reduce the magnitude
of the ``Hubbard''- like (on-site) interactions. 
Finally, for the JT - effect (which itself is nothing but another 
form of the Coulomb interaction) to come up there is no need to use
the Mn$^{3+}$ localized
states picture as it will be seen from what is done below.
The JT effect may persist even in the band model.

Therefore, in what follows, we adopt the band approach to describe
the ground state of LaMnO$_3$. This approach rationalizes the major
features of an insulating state in LaMnO$_3$ very well and merge  
into metallic state of ``doped'' manganites (see \cite{DGK}).

Experimentally it is known ~\cite{vmolnar} that LaMnO$_3$ 
transforms into the AF state below the $N{\acute e}el$ temperature, 
$T_N \simeq 140 K$.  Its magnetic structure corresponds to
the A - phase, i.e., consists of ferromagnetically aligned layers
which are coupled antiferromagntically along the one of main axes
(that axis is chosen as $c$-axis below). As for the lattice, the cubic 
cell becomes compressed along the same direction because the 
oxygen octahedra exercise the collective $JT$ ordering:
the octahedra are elongated along $x$- or $y$-direction and
are packed as it is shown schematically in Fig. 2. 
Correspondingly, the periodicity is doubled in the $(x,y)$ plane
by the JT - distortions, and is also doubled by the A - phase
 magnetic superstructure along the $z$-axis. Simple considerations 
will show that calculating the band spectrum for such a superstructure
and filling the bands at one electron per manganese site immediately
lead to the band insulator ground state.

We preface these calculations with a brief discussion of the role
of DE mechanism for a more simple model in which each ``manganese'' site has only
one orbital for mobile electrons (in addition to localized spin
${\bf S}_i$, of course). Instead of (\ref{eq:13}), one has now the
tight binding spectrum:
\begin{eqnarray}
\varepsilon ({\bf p}) = t (\cos{(p_x a)} + \cos{(p_y a)}
+ \cos{(p_z a)} ). 
\label{eq:19}
\end{eqnarray}
Imposing the Hund's term (\ref{dva}), $J_H \gg t$, the energy
of an electron in the ``half-metallic'' (fully polarized)
state, $\varepsilon_{hm}{({\bf p})}$, is: 
\begin{eqnarray}
\varepsilon({\bf p}) = - J_H S + \varepsilon ({\bf p})
\eqnum{19'}
\label{eq:19p}
\end{eqnarray}
and the electronic contribution to the total energy:
\begin{eqnarray}
E_{el}^{F} = \int{\varepsilon}_{hm}({\bf p}) n_{\bf p}
\frac{d^3 p}{(2 \pi \hbar)^3} \equiv -J_H S.
\label{eq:20}
\end{eqnarray}
(per cubic unit cell).

Rederiving Eq. (\ref{eq:16}) in the simplified case described above
for the A - phase, it is easy to find the electronic energy of
the A - phase (${\bf M} \simeq 0$ in $(\ref{eq:16}))$:
\begin{eqnarray}
E_{el}^{A} = - \int \sqrt{ (J_H S)^2 + t_{\bf p}^2} 
\frac{d^3 p}{(2 \pi \hbar)^3} \simeq - J_H S - 
\frac{t^2}{4~J_H~S}.
\eqnum{20'}
\label{eq:20p}
\end{eqnarray} 

From (\ref{eq:20}) and (\ref{eq:20p}) one may conclude first,
that terms of the order of $t^2/J_H$ in the electronic energy
make the AF - state (the A - phase) energetically more favorable.
That gain, being small, should be considered together with the
superexchange contributions, if any. Moreover, both 
Eqs. (\ref{eq:20}) and (\ref{eq:20p}) were obtained for the
``stoichiometric'' case of one electron per Mn site. The DE
mechanism in cooperation with large value of $J_H$ makes all electrons
have only one spin direction. Therefore, the bands in  
(\ref{eq:20}) and (\ref{eq:20p}) are fully occupied.
In {\it both} cases (ferromagnegnetic and antiferromagnetic states)
one would obtain the insulating phase for that model.

Returning to the more realistic situation of the $e_{2g}$-level,
the two bands,
and the A - type antiferromagnetic order, it follows again from
Eq. (\ref{eq:18}) that in the main approximation the problem
reduces to constructing the ground state of a single 
ferromagnetic layer. Unlike the previous case, a ``simple''
ferromagnetic state would be metallic because of the two bands:
\begin{eqnarray}
\varepsilon_{\pm}({\bf p}) = -{\mid}A{\mid}(c_x + c_y)
\pm {\mid}A{\mid} \sqrt{c_x^2 + c_y^2 - c_x c_y}
\eqnum{13'}
\label{eq:13p}
\end{eqnarray}
(since $|B|\ll{|A|}$ as was mentioned above,
the $B$-part in (\ref{eq:13}) was neglected) and one electron
per cell. The Brillouin zone will be only half-filled 
(all electrons are polarized). Superstructure imposed
by the $JT$ deformations, as shown in Fig. 1, suggests 
that since the Brillouin zone is now reduced by a factor of two,
the same number of electrons may fill up the reduced zone, producing 
an insulator at the appropriate energy spectrum. 
Experimentally, LaMnO$_3$ {\it is} an insulator and it is interesting
to find the electronic band spectrum in the presence of the
cooperative JT - distortions and then discuss the spectrum
properties together with the experimental findings.

Before we proceed further, let us make one remark. As it is
well known, two dimensional ferromagnetic state is never stable
being destroyed by spin fluctuations. Stabilization of the 
A - phase comes up due to the small terms in energy,
which are responsible for remnant interactions between layers.
Our estimates below for $\mid{A}\mid \sim 0.16 eV$ and
$J_H S \sim 1.5 eV$ would place the terms $\sim t^2/J_H$
on the scale of $\sim 100 K$ which agrees well with the low
value of the $N{\acute e}el$ temperature, $T_N \simeq 150 K$.

However, ``superexchange'' terms, $J{~\bf S}_i\cdot{~\bf S}_j$,
with $J$, being on the same scale as $\sim t^2/J_H$,  
may also become important. Role of these interaction
at the small ``electron'' doping (i.e. close to the 
CaMnO$_3$ end) has recently been discussed in terms of the 
band picture in \cite{Brink}.

To calculate the electronic spectrum of our model
in the presence of the ``antiferrodistorsive'' 
$JT$ collective deformations shown in Fig. 1,
let us note that the structure vector of the superlattice
is ${\bf q} = \frac{\pi}{a}(1,1,0)$. The superstructure
shows up in the JT - term (\ref{qbas}) and depends 
on which of the two modes $(Q_2, Q_3$ or even their 
superposition) of Eq. (\ref{qbas}) is chosen to correspond
 to the JT local deformations on each of the two sublattices.
Experimentally, the arrangement in Fig. 1 is close to the one in
which octahedra are elongated along each $x$- or $y$- axis preserving
the tetragonal symmetry in the perpendicular plane 
~\cite{vmolnar,ramirez} (For that mode the number of ``short legs''
$Mn-O$ equals four and the ``long'' ones equals two. These deformations
are also seen in some doped materials (e.g. see in \cite{Egami}).
Alternations of that specific type would be reflected in
expression (\ref{qbas}) as changes from $\theta=2\pi/3$ to
$\theta=-2\pi/3$ between  the two sublattices. That will lead
to the band's secular equation which can not be solved 
in the analytic form. However, one can consider
the contribution, which comes from the $Q_2$ mode only;
this deformation just changes
the sign on the adjacent sites of two sublattices, as also follows
from the Eq. (\ref{eq:8}) and Fig. 1. For that pattern one may
easily right down $\hat{H}_{JT}$ as:
\begin{eqnarray}
\hat{H}_{JT}({\bf q}) = -~\frac{gQ_0}{2}~\exp{(i{\bf q r})}
~\left ( \begin{array}{cc} 0 & 1 \\ 1 & 0 \end{array} \right ) =
-~\frac{gQ_0}{2}~\exp{(i{\bf q r})} \hat{\tau}_{x}
\label{eq:21}
\end{eqnarray} 
and obtain the spectrum by following the line of our
reasoning used above, when we were deriving the 
Eqs. (\ref{eq:16},\ref{eq:17}) ($\hat{\tau}_{x}$ is a
``pseudospin'' Pauli matrix defined on the basis (\ref{wfunct})).

Writing explicitly
\begin{eqnarray}
\begin{array}{c} 
\left [~\hat{\varepsilon} - \hat{t}({\bf p})~\right ]{\psi}_{\bf p} = 
- \frac{gQ_0}{2}~\hat{\tau}_x~\psi_{\bf p + q}, \\ \vspace{0.5mm}
\left [~\hat{\varepsilon} + \hat{t}({\bf p})~\right ]{\psi}_{\bf p + q} = 
- \frac{gQ_0}{2}~\hat{\tau}_x~\psi_{\bf p}
\end{array}
\label{eq:22}
\end{eqnarray}
(by definition $\hat{t}({\bf p+q}) = - \hat{t}({\bf p})$). Eliminating
${\psi}_{\bf p+q}$ from Eqs. (\ref{eq:22}), one obtains:
\begin{eqnarray}
\left [\varepsilon~\hat{\tau}_x + \hat{\tilde{t}}({\bf p}) 
\right ]\left [\varepsilon~\hat{\tau}_x - \hat{\tilde{t}}({\bf p})
\right ]
\left (\hat{\tau}_x~\psi_{\bf p} \right ) = 
\left (\frac{gQ_0}{2}\right )^2 \hat{\tau}_x~\psi_{\bf p}
\nonumber
\end{eqnarray}

The spectrum branches are then obtained from the zeroes of the
two-by-two determinant:
\begin{eqnarray}
det \left | \left ({\varepsilon}^2 - {\frac{gQ_0}{2}}^2 \right )
\hat{e} -  \hat{\tilde{t}}({\bf p})^2 + \varepsilon 
\left [\hat{\tilde{t}}({\bf p}),~\hat{\tau}_x\right ] \right | = 0,
\label{eq:23}
\end{eqnarray}
where $\hat{e}$ is the unit matrix, $\hat{\tilde{t}}({\bf p}) =       
\hat{t}({\bf p})~\hat{\tau}_{x}$, and 
\begin{eqnarray}
\left [\hat{\tilde{t}}({\bf p}),~\hat{\tau}_x\right ] = 
\hat{t}({\bf p}) - \hat{\tau}_x~\hat{t}({\bf p})\hat{\tau}_x.
\label{eq:24}
\end{eqnarray}
With $\hat{t}({\bf p})$ expressed in terms of the 
``pseudospin'' Pauli matrices:
\begin{eqnarray}
\hat{t}({\bf p}) = - f_{+}({\bf p})\hat{e} + 
\frac{1}{2} f_{+}({\bf p})\hat{\tau}_x + 
\frac{\sqrt{3}}{2} f_{-}({\bf p})\hat{\tau}_y,
\label{eq:25}
\end{eqnarray}
where $f_{+}({\bf p}) = \mid{A}\mid{(c_x + c_y)}, 
f_{-}({\bf p}) = \mid{A}\mid{(c_x - c_y)},$
after simple calculations, determinant 
(\ref{eq:23}) transforms to the form:
\begin{eqnarray}
\left |
\begin{array}{cc}
\left [{\varepsilon}^2 - (gQ_0/2)^2 - \frac{5}{4} f_{+}^2
+ \frac{3}{4} f_{-}^2 \right ] + 
i\frac{\sqrt{3}}{2} f_{+} f_{-} & f_{+}^2 - i\varepsilon\sqrt{3} f_{-} \\
f_{+}^2 + i\varepsilon\sqrt{3} f_{-} & 
\left [{\varepsilon}^2 - (gQ_0/2)^2 - \frac{5}{4} f_{+}^2
+ \frac{3}{4} f_{-}^2 \right ] - 
i\frac{\sqrt{3}}{2} f_{+} f_{-} \\
\end{array}
\right | = 0.
\nonumber
\end{eqnarray}
The resulting bi-quadratic equation produces the following
four branches, ${\varepsilon}_i({\bf p}) (i=1..4)$. Each of
these four branches is determined in the reduced Brillouin zone:
\begin{eqnarray}
{\varepsilon}_{1,2}({\bf p})&=& 
\left\{ (gQ_0/2)^2 + \frac{5}{4}f_{+}^2 + \frac{3}{4}f_{-}^2 \pm 
\sqrt{3 f_{-}^2\left [(gQ_0/2)^2 + f_{+}^2\right ] + 
f_{+}^4}\right\}^{1/2}\nonumber \\
{\varepsilon}_{3,4}({\bf p})&=&
\left\{ (gQ_0/2)^2 + 5/4 f_{+}^2 + 3/4 f_{-}^2 \pm 
\sqrt{3 f_{-}^2\left [(gQ_0/2)^2 + f_{+}^2\right ] + 
f_{+}^4}\right\}^{1/2}.
\label{eq:26}
\end{eqnarray}
At large enough ${\mid}{gQ_0/2}\mid$ the branches
${\varepsilon}_{1,2}({\bf p})$ are not crossing two
other branches, ${\varepsilon}_{3,4}({\bf p})$. Filling
them up by two polarized electrons per doubled unit cell,
completes the proof that, indeed, insulating LaMnO$_3$ may
be considered as {\it band} insulator.

For example, two sets of the spectrum branches (\ref{eq:26}),  
${\varepsilon}_{1,2}({\bf p})$ and 
${\varepsilon}_{3,4}({\bf p})$ begin to overlap
for ${\varepsilon}_{j=1}({\bf p})$ and
${\varepsilon}_{i=3}({\bf p})$ at ${p_x = p_y = \pi/2}$.
The overlap is direct 
which imposes some limit on the value of the $JT$ mode
which makes LaMnO$_3$ to be an insulator:
\begin{eqnarray}
\mid{gQ_0}\mid > 0.1 \mid{A}\mid.
\label{eq:27}
\end{eqnarray}

The optical gap, hence, corresponds to excitation of an 
electron from the ${\varepsilon}_{j=1(2)}({\bf p})$ band
into the ${\varepsilon}_{i=3(4)}({\bf p})$.

The optical spectra would, in principle, allow to access the gap directly.
The detailed analysis of the optical absorption in the pure
LaMnO$_3$ may be further complicated by the fact that formation of an
electron-hole pair by light, strictly speaking, should not lead 
to creation of individual electron- and hole-like excitations
due to inevitable polaronic effects in the {\it ionic} crystal LaMnO$_3$.
Since the experimental level of current optical studies 
(see \cite{optic}) in LaMnO$_3$ can not address
the issue yet, we shall not dwell upon further theoretical
analysis.

To complete this Section, one more simple comment may be helpful.
Namely, while in the case of a single ion with one electron 
occupies the {\it local} degenerate $e_{2g}$-level, it inevitably
leads to the local instability caused by the linear $JT$ term
(\ref{JT}), the ferromagnetic state with the band spectrum 
(\ref{eq:13}), filled up to some level, would remain stable
with respect to {\it small} enough JT distortions.
There is a {\it\bf threshold value} for the magnitude of the
JT - terms, before the distorted JT state may set in.       
This threshold is determined by the electronic kinetic energy
gain and the elastic lattice energy. In case of LaMnO$_3$ 
the existence of the cooperative JT deformations is confirmed
experimentally.
\section{Doped manganites. Percolation}
With the substitutional doping, A$_{1-x}$B$_{x}$MnO$_3$, say in,
La$_{1-x}$Sr$_{x}$MnO$_3$ by Sr, the materials initially show no metallic 
behavior with resistivity increasing by many orders of magnitude
at $T => 0$, In metals, one would expect that resistivity behaves
like
\begin{eqnarray}
\rho(T)&=&\rho_0 + A T^{\alpha},
\label{eq:29}
\end{eqnarray}
where $\rho_0$, the residual resistivity, is due to the
structure defects or impurities, while the $T$-dependence
comes up from scattering on thermal phonons or
from electron-electron interactions. In the second case,
the electronic relaxation rate $1/\tau_{ee} \sim T^2/E_F$, 
while the phonon mechanism prevails at 
$1/\tau_{ph} \sim T^3/{\theta}^2 > 1/\tau_{ee}$, i.e. at
$T > \theta (\theta/E_F)$, where $\theta$ is the characteristic
phonon frequency.
The latter in manganites is about four hundred degrees Kelvin
while, as we will see, for $E_F$ in manganites one gets the scale
$E_F \sim 0.1 eV$. Hence, the electron mechanism might become
important up to rather high temperatures. However, at elevated
temperatures a lot of new effects, related to the colossal 
magnetoresistance phenomenon, start to play the dominating role.

To understand the nature of the ground state of pure and doped 
manganites, we need,first of all, to concentrate our attention
on the low temperature properties. It has been shown first
in \cite{Urushibara} that low $T$ resistivity changes its character
with doping, so that the metallic behavior (\ref{eq:29})
sets in at $x=x_{cr}\approx{0.16-0.17}$ \cite{Okuda,Okimoto}. 
Interestingly enough, the onset of conducting regime coincides 
with the onset of the low temperature ferromagnetism, 
strongly indicating in favor of DE mechanism for the latter.

The value of $\rho_0$ may vary, even for nominally the same
material and composition, implying that quality of samples may
need further improvement. At the same time it is clear, that
in all cases the change from insulating to metallic behavior 
takes place when $x$ is close to the critical concentration, 
$x_{cr}$. 

The origin of that threshold and its value have been
first understood in \cite{LevandKresin,Levufn} in terms of percolation
theory (for review on percolation theory see e.g. the book 
\cite{Efros}). The latter considers any process
which, roughly speaking, corresponds to some exchange
between two adjacent local sites. A material, 
A$_{1-x}$B$_{x}$MnO$_3$, is commonly prepared by various methods
at high temperatures. As the result, position of atom $B$,
which is a substitution for a parent atom, is compelely random.
Divalent atom B, locally creates a ``hole'' on adjacent
Mn - sites which could serve as a charge carrier if there were
no long range Coulomb forces in the dielectric phase, which
keep the ``hole'' close to the negative charge at B$^{-}$.
When concentration is small, average distances between $B$ atoms
are large, which makes holes remain isolated. In the theory
of percolation one may look for the concentration at which 
the nearest neighbor B - atoms start to form infinite clasters 
piercing the whole crystal. For the B - atoms on the cubic 
sites it is known as the ``site'' problem: on the simple 
cubic lattice it gives
the critical concentration value at $\sim 0.31$. However, 
this is not exactly our case. For the doped holes, gathering on 
few manganese sites around an atom B, charge transfer 
takes place only along Mn-O-Mn bonds. Therefore the picture
of a critical cluster, constructed from the B$^{-}$-ions
must be corrected: such cluster would already have a finite
``thickness'' due to the holes spread over surrounding Mn - sites.
Numerical studies on the percolative models \cite{Scher} have shown
that this circumstance (i.e. the presence of the scale of a
few lattice constants) strongly decreases the value for the critical
concentration, rapidly converging to its value for the homogeneous
problem, $x_{cr} \sim 0.16$. The phenomenon of percolation in 
disordered media is akin to the other critical phenomena:
it bears a singular behavior in the vicinity of $x_{cr}$,
$\mid{x-x_{cr}}\mid~\ll x_{cr}$. Thus, it is expected that 
conductivity above $x_{cr}$ or $\rho_{0}^{-1}(x)$ in (\ref{eq:21})
should behave as $\sigma{(x)}{~\propto}(x-x_{cr})^{\gamma}$,
where $\gamma$ is a critical index. There were not many
attempts to verify singular behavior in conductivity.
The data \cite{Urushibara1,Urushibara} are consistent with
the index $\gamma \sim 0.5-0.6$ \cite{Egami}.

The above arguments have been given at $T=0$. One may try to apply
similar ideas to the CMR phenomena at $T=T_c$ (see, e.g., in 
\cite{LevandKresin,Levufn}). 
Indeed, if resistivity, $\rho{(T>T_c)}$ 
 is large enough, as it usually is, one may approximately
take the {\it conductivity}: $\sigma{(T>T_c)} = 
{\rho}^{-1}(T>T_c)~\simeq 0$. The fast increase in 
$\sigma{(T)}$ at $T<T_c$ is then expected to correspond
$\sigma{(T<T_c)}~\propto{(T_c - T)^{\overline{\gamma}}}$.
This is approximately true for La$_{0.8}$Sr$_{0.2}$MnO$_3$ near
$T_c$ with $\overline{\gamma}~\sim{0.6}$ \cite{Ghosh}.

The correctness of the percolation theory views may be also
verified by independent measurements of $\sigma{(T, x)}$ and
magnetization, $M(T,x)$, or $D_{stiff}(T,x)$ - the so-called 
``spin-stiffness'' which determines the spectrum of the
long-wave magnons:
\begin{eqnarray}
\omega{({\bf k})} = D_{stiff} {\bf k}^2
\label{eq:30}
\end{eqnarray}
($D$ itself is proportional to $M$).
The Kirkpatrick's relation gives:
\begin{eqnarray}
\sigma{\propto} D M.
\label{eq:31}
\end{eqnarray}
Eq. (\ref{eq:31}) may be verified at $x$ close to $x_{cr}$ 
or at $T$ near $T_c$.

For percolation in the continuous phases description the physical picture
is rather well understood. One of the helpful observations is that
for the three-dimensional problem there is a concentration range
in which percolation (infinite clusters) may take place
simultaneously for two phases (insulating and conducting, for
instance) though taking into account the surface tension 
at the boundary between two phases seems to impose limitations
on relative thickness of the phases.

In case of manganites, understanding on the microscopic level
remains far from being complete, though the concept 
is by all means correct. In the above, the mixture of ``two''
phases, depending on the concentration, $x$, may look as
intervened tiny ``islands'' and ``layers'' of different 
``phases'', whose thickness even difficult to access quantitatively.
 Well below and well above the threshold concentration $x_{cr}$,
one may imagine each corresponding phase as a bulk formation
into which the second phase is sparsely embedded. If there is
a spill-over of charge carriers between two phases, it is
the electro-neutrality condition which regulates the tiny domain sizes.

\section{Physical properties in two-band model}

As it was explained in the previous Section, the
percolation being a critical phenomenon, may cede soon
to the onset of the homogeneous ferromagnetic phase,
when screening becomes effective. In fact, some good
samples of La$_{1-x}$Sr$_{x}$MnO$_3$ show low temperature 
resistivity in the range $10^{-4}-10^{-5} (\Omega-cm)$
~\cite{Quijada}. Interest in applications of
CMR was the main reason why the properties of manganites
with $x~\sim 0.3$ have been elaborated rather carefully
(temperature of the resistivity peak, $T_p$, reaches its
maximum at 300 K for concentrations in this range).
There are still indications \cite{Egami} that remnants of
the second phase may still persist at $x~\sim 0.3$.
However, below we will analyze available low temperature experimental
data for doped manganites from the band approach. Our
conclusions will be that the band model (or Fermi liquid)
approach seems to be valid at low temperatures, especially
for the high quality samples. Data are hindered by sample's
quality with local inhomogeneities or the second phase inclusions
forming the scattering centers. For a number of compositions, 
depending on tolerance factor, conductivity is close to
its value in the mobility edge regime. The two band model,
thus, turns out to be a good starting point even at given
doping level, providing together with the interpretation 
of the parent material, LaMnO$_3$, an avenue for unifying 
theoretical understanding of the manganites properties.

Making use of the spectrum (\ref{eq:13}), 
we calculate the concentration dependence of the Fermi-level,
$E_F(x)$, the density of states (DOS), $\nu(x)$, the spin stiffness,
$D(x)$, and the whole magnon spectrum, $\omega({\bf k},x)$, and the
conductivity, $\sigma(\omega, x)$, both the Drude and the optical
(interband) components.  Theoretical results depend only on the 
single hopping integral, $|A|$.

We start with the calculations of the spin wave spectrum.
Let us write the  
deviations from the average spin, $\langle S_z\rangle$ for the localized
$t_{2g}$ - spins $({\bf s}={\bf S}-\langle S_z\rangle)$ as:
\begin{eqnarray}
{~s}^+({\bf q})=(2\langle S_z\rangle)^{1/2}&&\hat{b}({\bf q}),
{~s}^-({\bf q})=(2\langle S_z\rangle)^{1/2}\hat{b}^+({\bf q}),
\nonumber\\
&&{~s}_z({\bf q})=(\hat{b}^+\hat{b})_{\bf q},
\label{eq:32}
\end{eqnarray}
($\hat{b}^+, \hat{b}$ - the magnon's operators).  The first ($\delta
E_1$) and second ($\delta E_2$) order corrections to the
ground state are calculated as perturbations in:
\begin{equation}
\hat{V}=-J_H\sum_i{\bf s}_i(\hat{a}_i^+\mbox{\boldmath
$\sigma$}\hat{a}_i)=-J_H\sum_i{\bf s}_i{\bf n}_i.
\label{eq:33}
\end{equation}

For $\delta E_2$, the matrix elements in (\ref{eq:33}) are
of the form:
\begin{eqnarray}
V_{{\bf p}-{\bf k},\uparrow; 
{\bf p}\downarrow}^{l,l'}&=&-J_H\langle\uparrow
|s^{\pm}({\bf k})|\downarrow\rangle\nonumber\times\\
&&\times\left(\alpha_{{\bf p}-{\bf
k}}^l\alpha_{\bf p}^{\ast 
l'}+\beta_{{\bf p}-{\bf k}}^l\beta_{\bf p}^{\ast l'}\right),
\label{eq:34}
\end{eqnarray}
(the coefficients $\alpha_{\bf p}^l, \beta_{\bf p}^l$ are defined below). 
As for $\delta E_1$, its only role is to secure the proper behavior of
the magnon spectrum at $k\rightarrow 0$).  
One obtains:
\begin{eqnarray}
\delta E_2=&&2 J_H^2\sum_{\bf k}\langle S_z\rangle
\hat{b}^+({\bf k}) \hat{b} ({\bf k})\times\nonumber\\
&&\times\sum_{l,{\bf p}}\left(
\sum_{l'}\frac{|\alpha_{\bf p}^l\alpha_{{\bf p}+{\bf k}}^{\ast
l'}+\beta_{\bf p}^l\beta_{{\bf p}+{\bf k}}^{\ast
l'}|^2}{E_{\uparrow}^l({\bf p})-E_{\downarrow}^{l'}({\bf p}+{\bf k})}\right),
\label{eq:35}
\end{eqnarray}
where
\begin{eqnarray}
E_{\uparrow, \downarrow}^{l, l^{\prime}}({\bf p})&=&
\mp J_H\langle S_z\rangle
+\varepsilon_{l, l^{\prime}}({\bf p}).
\label{eq:36}
\end{eqnarray}
Both sums in (\ref{eq:35}) run over 
$l,l^{\prime}=\pm$ (Eq.(~\ref{eq:13})).  
The summation over $l$ and ${\bf p}$ is limited by the occupied states
$(\uparrow)$ only.  The coefficients $(\alpha_{\bf p}^l, \beta_{\bf
p}^l)$ above are for the Bloch's states on the basis (\ref{wfunct}):
\begin{equation}
\alpha_{\bf p}^{l,l'}=\left(\Sigma_{12}/2|\Sigma_{12}|\right)^{1/2},
~\beta_{\bf p}^{l,l'}=\pm\left(\Sigma_{21}/2|\Sigma_{12}|\right)^{1/2},
\label{eq:37}
\end{equation}
(here $\Sigma_{12}, \Sigma_{21}$ are the off-diagonal elements of the 
hopping matrix $\hat{t}$({\bf p}) 
in (\ref{eq:11}) on this basis).

Assuming in (\ref{eq:35}, \ref{eq:36}) 
(recall that $J_H\gg
|A|$ ($J_H S_z$ is of the order 
of 1.5 eV  \cite{vmolnar} , while  
estimates below produce
for $|A| \sim 0.1$ eV),
expanding (\ref{eq:35}) in $|A|/J_H$ 
would get a series of the Heisenberg spin Hamiltonians
accounting for interactions with the increasing number of neighbors.
(For a single band it was first noticed in  \cite{Furukawa}; in this
paper we are using the realistic two bands picture).  
After a somewhat tedious, but straightforward calculation,   
 the first order term in $\mid A\mid$ from (\ref{eq:35}) is $(\langle
S_z\rangle =3/2)$:
\begin{equation}
\hbar\omega({\bf k})=|A|(3-c_x-c_y-c_z)D(x)/3
\label{eq:38}
\end{equation}
and $D(x)\equiv D(E(x))$ is given by the integral:
\begin{eqnarray}
\int\frac{d^3{\bf p}}{(2\pi)^3}\left[\, 
\sum_{(+,-)}\theta(E-\varepsilon_i({\bf p}))
\left\{
1\pm\frac{2c_x-c_y-c_z}{2 R({\bf p})}\right\}
\right ],
\nonumber
\end{eqnarray}
(here $E$ is in units of $|A|, p_i\equiv a p_i$).  
Quantum fluctuations may change  the {$\bf k$} -- dependence 
in Eq. (\ref{eq:38}).

Let us turn now to the calculation
of conductivity, $\sigma_{ij}(\omega, x)$, which is
described by:
\begin{eqnarray} 
{\sigma}_{ij} = -\frac{e^2{\hbar}^2}{V\omega}\sum\limits_{\bf k,k'}^{}
f_0({\bf k})\left [1 - f_0({\bf k'})\right ]
\langle\psi({\bf k})\mid\hat{v}_i\mid\psi({\bf k'})\rangle
\langle\psi({\bf k'})\mid\hat{v}_j\mid\psi({\bf k})\rangle
\times\nonumber\\
\left [\delta(\epsilon({\bf k'}) - \epsilon({\bf k}) - \omega) - 
\delta(\epsilon({\bf k'}) - \epsilon({\bf k}) + \omega)\right ],
\end{eqnarray}
where $f_0({\bf k})$ is the Fermi distribution function,
${\bf k}$ is a quasi momentum, $\hat{v}_i$ is a velocity
operator. For the cubic crystal $\sigma_{ij}=\sigma_{ji}$. 

We determine first the velocity operator,
$\hat{\bf \hat{v}}=\hat{\bf \dot{r}}$ (see \cite{Pit}):
\begin{equation}
\hat{\bf v}({\bf k})=\frac{1}{\hbar}
\frac{\partial\varepsilon_l({\bf k})}{\partial
{\bf k}}+\frac{i}{\hbar}[\varepsilon_l({\bf k})-
\varepsilon_{l'}({\bf k})]\,\langle
l{\bf k}|\hat{\mbox{\boldmath $\Omega$}}|l'{\bf k}\rangle.
\label{eq:39}
\end{equation}
The off-diagonal operator $\hat{\mbox{\boldmath $\Omega$}}$ is
defined by the relation:
\begin{equation}
\langle l{\bf k}|\hat{\mbox{\boldmath $\Omega$}} |l'{\bf k}\rangle =i\int
u_{\bf k}^{\ast l'}({\bf r})\frac{\partial
u_{\bf k}^l}{\partial {\bf k}} d^3{\bf r}
\label{eq:40}
\end{equation}
and $u_{\bf k}^l({\bf r})$, the periodic Bloch functions on the basis
of Eq. (\ref{wfunct}) are:
\begin{eqnarray}
u_{\bf k}^l({\bf r})&=&\frac{1}{\sqrt{N}}\sum_n\exp
[i{\bf k}({\bf a}n-{\bf r})]\times\nonumber \\
&& \times ~\left\{\alpha_{\bf k}^l\phi_1({\bf r}-n{\bf
a})+\beta_{\bf k}^l\phi_2({\bf r}-n{\bf a})\right\}.
\label{eq:41}
\end{eqnarray}
With the one-site integrals only in (\ref{eq:40}) and 
Eqs. (\ref{eq:37}):
\begin{equation}
\langle l{\bf k}|\hat{\mbox{\boldmath $\Omega$}}|l'{\bf k}\rangle
=i\frac{a}{\hbar}\frac{\sqrt{3}}{4}\frac{(-\sin
k_x)(c_y-c_z)}{|t_{12}|^2}.
\label{eq:43}
\end{equation}
Matrix elements in (\ref{eq:39}, \ref{eq:43}) produce transitions 
from {\it occupied} parts of the $\varepsilon_+({\bf p})$-band 
into {\it empty} states
in the $\varepsilon_-({\bf p})$-band.

With all the above, we arrive to the Drude (intraband)
contribution which in the clean limit is:
\begin{eqnarray}
\sigma_{Drude}(\omega, x)&=&2\pi\frac{e^2 |A|}{3a\hbar^2}\delta(\omega)
I_{Dr}(x),\label{eq:44} \\
I_{Dr}(x)&=&\frac{1}{2(2\pi)^3}\sum_l\int dS_{\bf p}^l|\mbox{\boldmath
$\nabla$}_{\bf p}\varepsilon({\bf p})|,
\label{eq:45}
\end{eqnarray}
(the integral in $I_{Dr}(x)$ is over the Fermi surfaces).

The ``optical'' (interband) contribution is 
\begin{eqnarray}
\sigma_{opt}&(&\omega,x)=\frac{3\pi e^2}{a\hbar}\frac{1}{\tilde{\omega}_0^3}
\int\frac{d^3{\bf p}}{(2\pi)^3}\sin^2p_x(c_y-c_x)^2 
 \times \nonumber \\ 
&& \times ~n(\varepsilon_+({\bf p}))[\,1-n(\varepsilon_-({\bf p}))]  
\cdot \delta\left(
\tilde{\omega}-2 R({\bf p})\right),
\label{eq:46}
\end{eqnarray}
where $|A|\tilde{\omega}=\hbar\omega$ (Eqs. (\ref{eq:45}) and 
(\ref{eq:46}) agree with the similar calculations 
of \cite{Takashi}).
For the low temperature spectral
weight
\begin{eqnarray}
N_{eff}={{2m}\over{\pi e^{2}}} a^3 
\int\limits_{0}^{\infty}\sigma(\omega) d\omega
\label{eq:47}
\end{eqnarray}
one obtains both the Drude and the interband contributions,
respectively:
\begin{eqnarray}
N_{eff}^{Drude} = \frac{ma^2}{3\hbar^2}|A|I_{Dr}(x),
~N_{eff}^{opt} = \frac{ma^2}{\hbar^2}|A|\frac{3}{4}I_{opt}(x),
\label{eq:48}
\end{eqnarray}
(with $I_{opt}(x)$ directly obtained from (\ref{eq:46})).
In Fig. 2 we plotted our results for the Fermi level, 
$E_{F}(x) = |A|E(x), 
\tilde{\nu}(x)={\nu}(x)|A|, 
D(x)$ and $I_{Dr}(x)$ (the shaded area
shows schematically the concentration range for a percolative
regime).
In Fig. 3 evolution of the Fermi surface with $x$ is shown
(Fermi surfaces in Fig. 3a, 3b, 3c are for concentrations 
$x=0.2, 0.3, 0.5$ respectively. An interesting fact is that
$x=0.3$ is the concentration at which a ``neck'' 
develops at the zone boundary of the Fermi surface. 
In other words, concentration $x=0.3$ is the point
of ``2.5'' - Lifshitz transition at which the change
in the Fermi surface topology occurs.

\section{The Model and Experiment. Discussion}

In the discussion below we restrict ourselves 
mainly by the ``isotropic'' materials, i.e. 
by the range of concentrations $x\le 0.4$.
Around $x\sim 0.5$ new phenomena start to develop
in doped manganites, such as charge ordering \cite{Cheong,Tomioka}, 
$A$- phase in Pr$_{0.5}$Sr$_{0.5}$MnO$_3$ and 
Nd$_{0.45}$Sr$_{0.55}$MnO$_3$ \cite{Kawano,Akimoto}, 
structural transitions caused by 
magnetic field in Pr$_{0.5}$Sr$_{0.5}$MnO$_3$, 
Nd$_{0.5}$Sr$_{0.5}$MnO$_3$ \cite{Mahendiran,Tokura,Tokunaga}
and ``spin valve'' effect in
Nd$_{0.46}$Sr$_{0.54}$MnO$_3$ \cite{Kuwahara}. 
Though some preparatory steps
have been made to address these issues, e.g. Eqs. 
(\ref{eq:14}, \ref{eq:14p}) in Sec. III, the purpose 
of this paper is mainly to access the limits of
applicability of the band structure theory above, and its
parameters. For these reasons the concentration range
$x\sim 0.3$ is chosen where more data are available.
A word of caution still would be helpful. Although the 
quality of materials has been dramatically improved 
for the past couple of years and many results are becoming
quite reproducible, systematic studies of the same phenomenon
on the same materials, which would have been performed in different
laboratories, are rather rare.
It is also not well known how data, obtained on single crystals
may differ from the ones obtained in crystalline films.

The spectrum (\ref{eq:13}) and the results above, such as
(\ref{eq:38}, \ref{eq:44}-\ref{eq:48}) include the single
energy scale, $\mid{A}\mid$. We choose to determine this
parameter from the measured spin stiffness coefficient in
(\ref{eq:38}):
\begin{eqnarray}
\hbar{\omega}({\bf k})&=&\frac{({\bf ka})^2}{6} \mid{A}\mid D(x),
\eqnum{38'}
\label{eq:38p}
\end{eqnarray}
i.e.
\begin{eqnarray}
D_{stiff}&=&\frac{a^2}{6}\mid{A}\mid D(x).
\nonumber
\end{eqnarray}
Being a long-wave characteristics, a latter must be less sensitive
to defects of sample's quality. We use $a\simeq 3.86 \AA, 
D(x\simeq 0.3)\simeq 0.45$ (according to our results in Fig. 2) 
and the data, collected in ~\cite{Fernandez} (Table 1). The
results for $\mid{A}\mid$ show that the bandwidths for different
materials, $W=6\mid{A}\mid$, do not vary significantly 
($W\simeq 0.7-1.0 eV$). This is in favor of the fact that 
the tolerance factor itself which is responsible for variations in
the $Mn-O-Mn$ bonds angle for the different materials, is {\it not}
of much importance, as it has already been suggested for
the bandwidth in 
\cite{DGK,Egami}.
As for the concentration dependence, $D(x)$ in Fig. 2,
it reflects the fact that number of electrons and, hence, 
the saturation moment $M(x)$ decreases with the increase of $x$.
There are measurements of $D_{stiff}(x)$ \cite{Moudden,Hirota} which 
show some pronounced dependence on $x< 0.2$ and saturation
at $x=0.28, 0.3$ \cite{Moudden}. The decrease in $D(x)$, 
according to our results, shown in Fig. 2, at $x\sim 0.3$
is below the experimental accuracy in \cite{Moudden}. As for
measured dependence $D_{stiff}(x)$ for smaller concentrations,
here one may meet the range for the percolative behavior,
shown in our Fig. 2 by the shaded area. We will return 
to it later, although it is worth mentioning that data
\cite{Moudden,Hirota} do not agree with each other above $x=0.2$.

With $\mid{A}\mid\simeq 0.16 eV$ (La$_{1-x}$Sr$_{x}$MnO$_3$) 
one may find the thermodynamical parameters and 
the density of states, $\nu{(x)}=\tilde{\nu}(x)/\mid{A}\mid$. 
Correspondingly, the Sommerfeld coefficient, $\gamma$,
 for our spectrum (\ref{eq:13}) is
\begin{eqnarray} 
\gamma&=&{\pi}^2 \tilde{\nu}(x)/3\mid{A}\mid.
\label{eq:49}
\end{eqnarray}
An attempt to investigate properties of La$_{1-x}$Sr$_{x}$MnO$_3$ 
systematically has been done by Okuda, where special efforts
were applied to single out the electronic component into the
specific heat by subtracting both phonon and magnon contributions.
According to \cite{Okuda}, $\gamma\simeq{3.5} mJ/mole\cdot{K^2}$,
while $\tilde{\nu}(x)\simeq 0.45$ at $x=0.3$ from Fig. 2
produces $\gamma\simeq{6.2} mJ/mole\cdot{K^2}$. Note, 
however, that $\tilde{\nu}(x)$ has a kink at $x=0.3$. That kink
takes its origin from the fact that this concentration is the
point of the Lifshitz singularity, as it may be clearly seen from
the Fermi surface pattern in Fig. 3. The total magnon contribution
into specific heat is proportional to $T^{3/2}$. The proximity 
to the Lifshitz ``2.5'' transition results in appearance of the
same $T-$ dependence of the electronic specific heat. As the result,
the procedure, \cite{Okuda}, of extracting magnon $T^{3/2}$ terms 
becomes less transparent. It is worth mentioning that 
unambiguous determination of the electronic contribution is known
\cite{vmolnar} as a 
difficult task. The low $T$ electronic specific heat of a few other
manganite compounds has also been measured with $\gamma$ in
the $3-8 mJ/mole\cdot{K^2}$ range. Taking into account the possible
complications with the rapid energy dependence in the density of 
states close to $x=0.3$, the accuracy of Eq. (\ref{eq:49}) seems 
to us rather satisfactory.  

Important issue is a sensitivity of all data to disorder, which is
inevitably present at the substitutional doping. The concentrations
are not small by all means. The residual resistances in Eq. (\ref{eq:29})
obtained for nominally the same compositions, may vary significantly for 
different groups of compounds. So, it is clear, that to some extent,
the best values of $\rho_{0}(x)$ are still to be determined.
Meanwhile, it remains unclear whether $\rho_{0}(x)$ only characterizes
the sample's quality or there is an intrinsic component in the
residual resistance. In favor of the former suggestion tells the
results of recent experiments performed by 
\cite{Lofland1,Quijada,Lofland}.
New records have recently been reported for conductivity, 
measured in a few compounds.
Among them there are resistivity data for La$_{1-x}$Sr$_x$MnO$_3$ 
in crystalline films which give the value for $\rho_{0}(x)$ 
as low as $10^{-5} \Omega\cdot{cm}$ \cite{Quijada}. 
This is a typical metallic
conductivity range. Making a substitution in Eq. (\ref{eq:44})
for $\pi\delta{(\omega)} -> \tau/(1+(\omega\tau)^2)$, we obtain
in this case $\hbar/\tau\mid{A}\mid\sim 2\cdot{10^{-2}}$, 
i.e. for the inverse life time, 
$\hbar/\tau\sim 3\cdot{10^{-3}} eV
(\sim 30 K)$. At the same time the residual resistivity for other
materials studied in \cite{Quijada}, lie in the range 
$100\div{300} (\mu\Omega - cm)$, i.e. 
$\hbar/\tau\mid{A}\mid\sim 0.5$. These findings become more transparent
being expressed in terms of the mean free path. With the average
velocity of an electron on the Fermi surface
\begin{eqnarray}
\overline{v}=<{\bf v}^2>^{1/2} = (\mid{A}\mid{a}/\hbar)
(2 I_{Dr}(x)/\tilde{\nu}(x))^{1/2}, 
\label{eq:50}
\end{eqnarray}
the mean free path, $l=\overline{v}\tau$, is typically $\sim 3a$ for
materials with $\rho_{0}\sim 10^{-4} (\Omega\cdot{cm})$, while
in the best $LSMO$ samples it is around $80a$. Whether the 
values of $\rho_0$ mentioned above may be or may be not improved
by a more careful sample preparation process, remains to be
seen. In any case, for some current materials \cite{Quijada} 
the conductivity regime
lies close to the mobility edge. If the short mean free path in
these materials is an intrinsic feature, it may be related
to the local fluctuations in the tolerance factor (\ref{tol})
caused by the difference in ionic radii at the $Sr-$ substitution.
Indeed, with $<r_0>=0.14nm, <r_{Sr}>=0.129 nm, <r_{La}>=0.136 nm$ for
$Pr^{3+}$ and $Nd^{3+}$ one has $<r_{Pr}>=0.129 nm$ and
$<r_{Nd}>=0.124 nm$ \cite{vmolnar}(p. 194). As it has been emphasized 
already, our two--band approach is a simple way of the Fermi
liquid description, when interactions are not supposed 
to be remarkably strong. The strength of interactions is 
provided by the $T^2-$ term in resistivity, in which
electron interactions come from 
$\hbar/{{\tau}_{ee}}^{tr} = {\lambda}^{\prime}\left(\hbar/{\tau}_{ee}
\right)$, where ${\tau}_{ee}$ is the total quasiparticle 
relaxation time and ${\lambda}^{\prime}<1$ gives the fraction 
of Umklapp processes.
We use \cite{Gantmaher}:
\begin{eqnarray}
\hbar /\tau_{ee}\simeq \lambda{\pi}^3{\nu}(x)T^2 .
\label{eq:51}
\end{eqnarray}
In (\ref{eq:51}) $\lambda$ is a value of the interaction strength
in terms of $E_F$. Using \cite{Quijada,Lofland}, one obtains for LSMO:
${\lambda}{\lambda}^{\prime}\simeq 0.3$, typical of good metals.
As for the two other materials in \cite{Quijada}, the $T$-variations
of resistivity scale in the magnitude with their residual resistivity
and most probably are caused by defects \cite{Mahan}.

$T$- dependence in the optical conductivity, $\sigma(\omega)$,
attracted recently much attention  \cite{Okuda,Quijada,Boris} 
as a manifestation of changes in the conductivity mechanism 
from metallic to localized polarons at
elevated temperatures. We discuss only a few results pertinent
to the low $T$ band mechanisms. First note, that the
temperature dependence in $\sigma(\omega)$ at
$T<100$ K for $\omega$ around 1eV is most pronounced 
{\it below} 1eV (see Fig. 2 in \cite{Quijada} ).  
This agrees well with our estimates for the bandwidths,
$W \leq 1$eV.  As for a quantitative analysis, there
are problems of an experimental character.  
Assuming $N_{eff}(\omega)$  \cite{Quijada}  would give our
$N_{eff}$'s in Eq. (\ref{eq:48}) 
at $\omega\simeq 1$eV and that both the
Drude and optical contributions are approximately equal \cite{Takashi}, 
we obtain
$N_{eff}\sim 0.25$ which is reasonably close to the values 
in  \cite{Quijada} ,
lesser than $N_{eff}$ for the single crystal data,
La$_{0.67}$Ca$_{0.33}$MnO$_3$  \cite{Boris} , and a factor ten bigger than
$N_{eff}$ for La$_{0.7}$Sr$_{0.3}$MnO$_3$ 
in  \cite{optic} .  We believe that such
a difference originates from poor data for the optical $\sigma
(\omega)$ in the ``Drude-tail'' range.

Finally, the optical gap for pure LaMnO$_3$ was 
identified \cite{optic} at $\Delta\simeq 1.2$eV.
A rough estimate from the band insulator picture
(\ref{eq:26}) gives $gQ_0/2\sim 0.6 eV$,
i.e. the $JT-$ coupling is rather strong 
($\mid{A}\mid{\sim} 0.16 eV$).

We conclude the discussion by a few comments regarding the spin
wave spectrum.  In La$_{0.7}$Pb$_{0.3}$MnO$_3$ 
\cite{Perring} the spectrum fits
well (\ref{eq:38}). Eq.(\ref{eq:38}) follows from (\ref{eq:35}) 
at $A \ll J_H\langle S_z\rangle$, with quantum corrections neglected
(we have also neglected terms of the order of $t^2/J_H$).
Meanwhile, strong deviations from (\ref{eq:38}) have been
observed at $\xi\,>\,0.25$ along the (0,0,$\xi$)-direction 
in Pr$_{0.63}$Sr$_{0.37}$MnO$_3$  \cite{Hwang}. The spin stiffness
changes only slightly from material to material \cite{Perring},  
and other low temperature characteristics including strength of electron 
interactions, also seem not to vary much. 
Unlike \cite{Perring}, we suggest with \cite{Hwang} that 
differences in low-$T$ spin dynamics for
two materials do not correlate with their behavior at higher temperatures,
but with a tendency to a low temperature charge ordering or 
formation of the A - phase at
$x$ close to 0.5 (we will address this issue elsewhere).

Let us now return to the problem of how doped manganites
proceed from the insulating end, LaMnO$_3$, with the
increase of concentration, $x$. At $x<x_{cr}$ doped holes
remain localized in the vicinity of corresponding clusters.
If the clusters are small enough they might be called 
``ferromagnetic polarons'' but this is a semantics. In [deGennes, 1960]
it has been argued that such isolated microscopic ferromagnetic
inclusions do produce a macroscopic consequences, forming 
``canted'' ferromagnetic moment even in the $A-$ phase of parent
LaMnO$_3$. In other words, the effective radius of ferromagnetic
interactions is larger than short range coupling in the percolative
scenario for conductivity. The relation (\ref{eq:31}) based on the
assumption of equal effective radii both for conductivity 
hopping and ferromagnetic interactions, is, thus, not expected 
to be applicable near $x_{cr}=0.16$:
magnetic ordering and ferromagnetic component in the spin waves, 
may develope
earlier. The experimental results \cite{Moudden,Hirota}, indeed,
show the stiffness coefficient, $D_{stiff}(x)$ nonzero below $x_{cr}$
and starting to increase almost at $x<0.15$. However, the further
sharp increase above $x_{cr}=0.16$ with a saturation near
$x\simeq 0.3$, is the bright demonstration in favor of the notion
of percolation: the increase in $D_{stiff}(x)$ just follows increase in the
amount of the ferromagnetic component. Note, that $D_{stiff}(x)$ is proportional
to the value of magnetization, $M(x)$. In the band picture, 
$M(x)$ is $(4-x)\mu_{B}$. As it is known since the old results \cite{vmolnar}
that $M(x)$ actually increases at $x>0.1$. Interesting thing is that
$M(x)$ has dependence, close to $(4-x)\mu_{B}$, 
even at $x\sim 0.3-0.4$, indicating that admixture of ``insulating
phase'' still persists at these concentrations. 

Another proof comes from recent pulsed neutron experiments
\cite{Egami,Billinge}. These experiments probe the local
arrangement of the oxygen octahedra. With the random disorder, 
one may expect that the degeneracy of the $e_{2g}-$ terms
would also be lifted randomly, providing a distribution 
in the JT deformations of the oxygen octahedra. Meanwhile,
it has been shown in \cite{Egami} that the pair distribution function
displays the well pronounced peaks at the values of $Mn-O$ bond
which are characteristic for the parameters of the elongated octahedra
in the parent LaMnO$_3$. According to \cite{Egami}, in 
La$_{1-x}$Sr$_{x}$MnO$_3$ the presence of such insulating inclusions
is seen up to $x=0.35$.
The local neutron probe \cite{Egami,Billinge} can not resolve
the size and the pattern of the distorted areas. One possible view
may be to consider them as randomly positioned centers or the $JT$ 
``polarons'' with a scale $\sim 15-20 \AA$. However, we have to keep
in mind, that in the 3D percolative regime, the percolative 
paths (infinitely connected clusters) may coexist for both phases
simultaneously. The analysis of low temperature metallic properties
performed in the previous Section, seems to indicate that at least
for concentrations $x\sim 0.3$, the ferromagnetic phase component,
in the first approximation,
occupies most of the bulk with other phase embedded into it
and also seen as reasonably small scattering centers.

The Kirkpatrick relation (\ref{eq:31}) is expected to be approximately
correct in homogeneous media regime, near $T_c$, where onset of
ferromagnetism may be interpreted as the simultaneous onset of
a new (metallic) mechanism for conductivity. As it was pointed out
at the end of Section VI, Eq. (\ref{eq:31}) may be satisfied reasonably well.
At the same time, it has been already known \cite{Lofland}
that in La$_{0.8}$Sr$_{0.2}$MnO$_3$ the peak temperature, $T_p$, and the
Curie temperature, $T_c$, do not coincide ($T_p=$ and $T_c=$).
These experiments were mostly considered in the literature 
as a crossover \cite{Millis1} in which conductivity regime 
via the thermal hopping of polarons localized by the thermal 
lattice disorder cedes sharply to a metallic 
regime with a short mean free path.
In recent STM experiments \cite{Mydosh}
it has been successfully demonstrated that even around
$T_c$ the regime in La$_{0.7}$Ca$_{0.3}$MnO$_3$ is better described 
as a coexistence of the two percolative phases: metallic and
insulating ones. 
The characteristic scale for sizes of subphases having the shape
of smooth clouds are from tens to hundreds of nanometers.

Another experimental support comes from the M\"ossbauer spectroscopy
measurements \cite{Nath}. The strong paramagnetic signal has been
observed at $T > T_c$. Decrease in $T$ leads to the appearance
at $T < T_c$ of the strong signature of the ferromagnetic state,
namely the ``6 picks'' structure in the M\"ossbauer
signal. However, even at $T < T_c$ the paramagnetic signal
persists up to $T\simeq 20$ K, which is much lower then $T_c$.
These studies strongly support the exsistance of the two
phases and therefore the percolative scenario.
  
\section{Conclusions}
We have discussed above two major issues in doped manganites:
1. -- the concept of percolation, leading, as it was first predicted 
in \cite{LevandKresin}, to the new understanding of the phenomena related to the
phase separation, which takes place in these materials at low
enough concentrations;2. -- the possibility to interpret some
low temperature data in the habitual band (or Fermi liquid)
picture. The latter makes possible to suggest an unifying
view which circumvent the often raised concerns regarding the role
of strong interactions, importance of which is commonly expected in
the transition metal oxides \cite{Godunov}. The Fermi liquid approach
is a selfconsistent theory, which of course, may be broken
at strong interactions. The only way to justify the Fermi liquid approach
is to see whether the experimental results do fit the theory
predictions or not. As it was shown above, the theory works reasonably
well at $x=0$ and at $x\sim 0.3$ for $La_{1-x}Sr_{x}MnO_3$, 
probably because the $DE-$ mechanism results in rather effective
de-localization while the JT - effect partially takes care of 
the Coulomb interaction. In the intermediate regime with respect to $x$,
the tendency towards the phase segregation turns out to have a percolative
character and, therefore, theoretical analysis becomes much more
complicated.

The question whether the samples quality is an intrinsic property
or may be further improved, seems to be of principle importance.
Experience with LSMO materials \cite{Quijada} implies that attempts made to improve
the crystals quality are worth of trying. Already the LSMO samples
with $1/\tau\sim 20 K$ are encouraging, being reasonably close to 
the possibility of studying metallic manganites by methods traditional
to the normal physics of metals, 
including the possibility of the de Haas - van Alphen studies
in fields of the order of $30 - 40$ Tesla.

\begin{acknowledgments}
L.P.G. and V.Z.K. gratefully acknowledge H.D. Drew,
T. Egami, J. Lynn, S. von Molnar for
numerous stimulating discussions.
The work was supported (L.P.G. and M.O.D.) by the National High Magnetic
Field Laboratory through NSF Cooperative Agreement \# DMR-9527035 and the
State of Florida.
\end{acknowledgments}

\newpage
{\bf Figure Captions} \\
{\bf Fig. 1} In-plane staggered distortions inside the ferromagnetic
layer. Solid segmenets represent the elongations of octahedra.
The new unit cell is shown as the dashed square.\\
{\bf Fig. 2}  The Fermi level, E(x), DOS, $\tilde{\nu}(x)$, the spin
stiffness coefficient D(x) and the Drude conductivity, 
$I_{Dr} (x)$, plotted as a function of concentration, x, 
for the spectrum, given in \cite{DGK}.
The shaded area shows the percolative
regime where Eqs. (37-46) are not applicable.\\    
{\bf Fig. 3} The Fermi surface a) x=0.2; b) x=0.3; c) x=0.5.\\
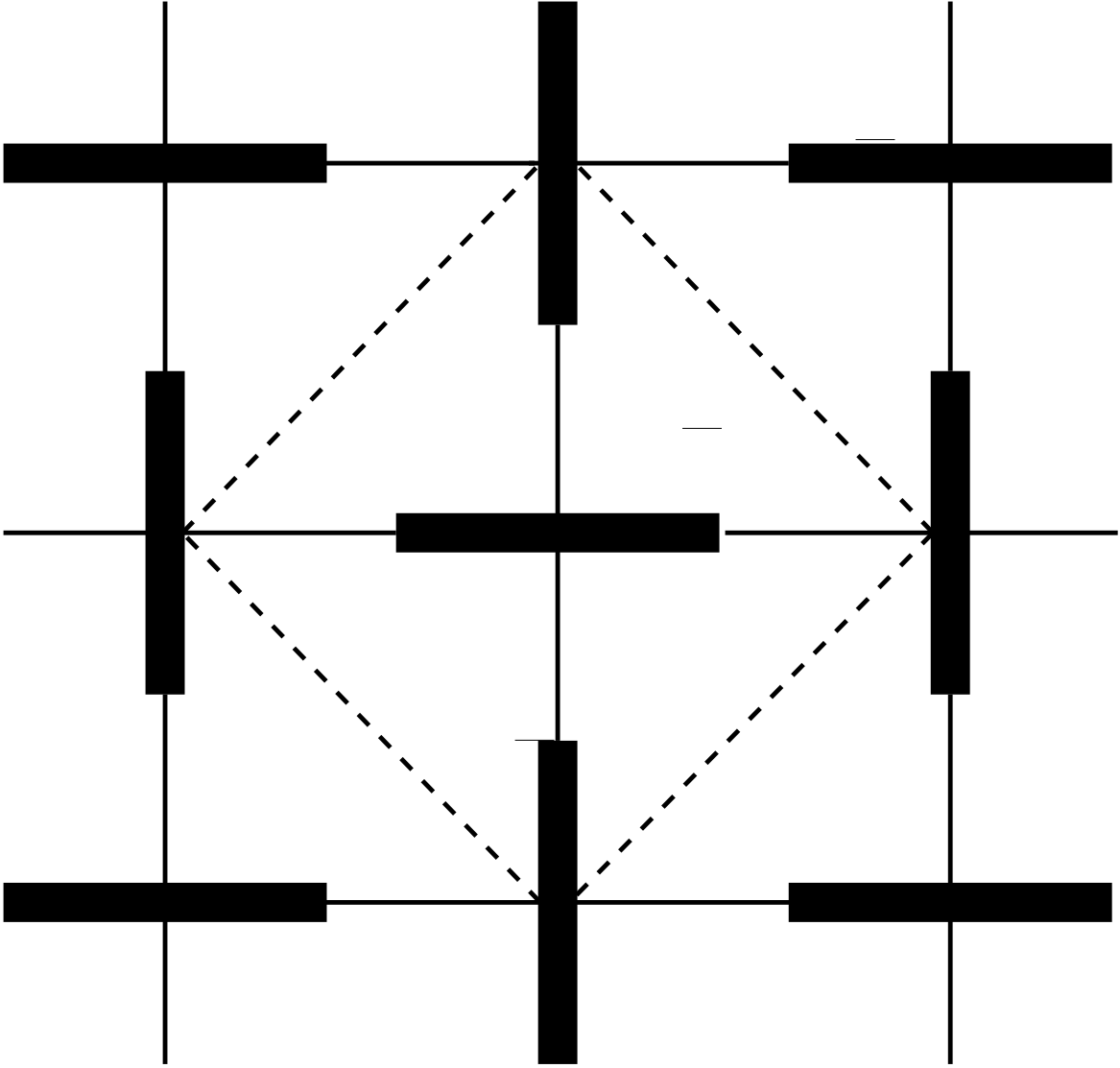
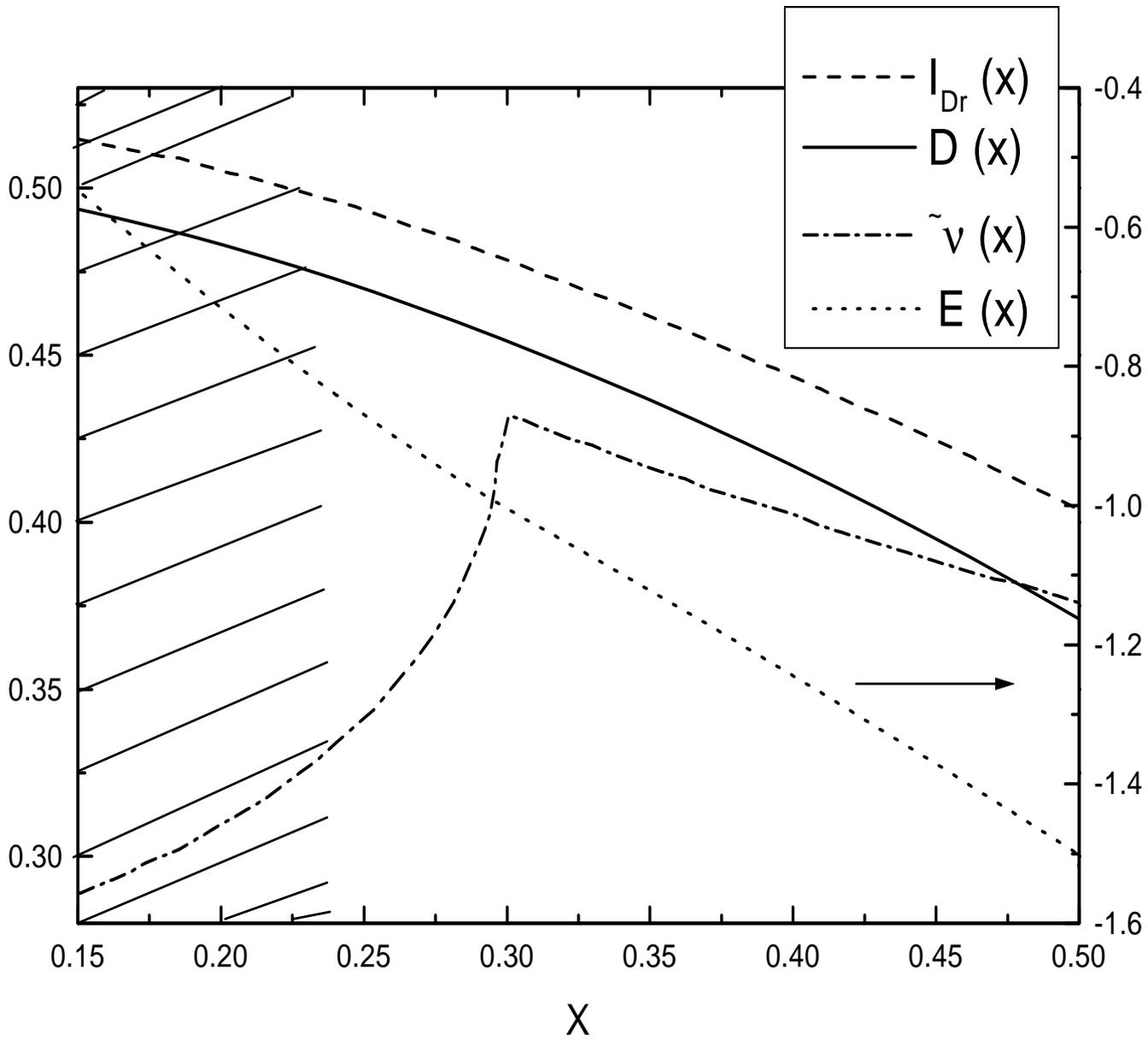
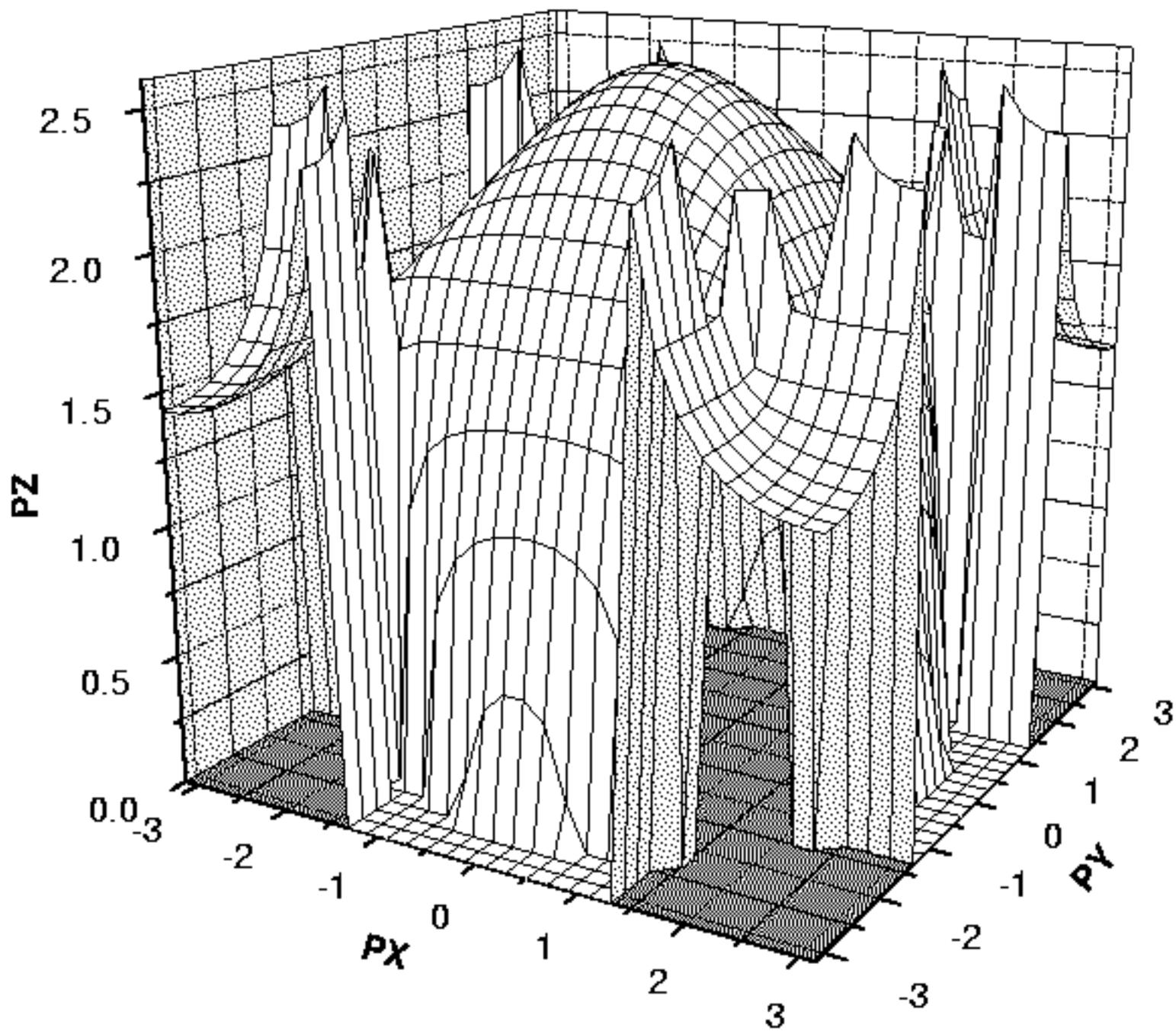
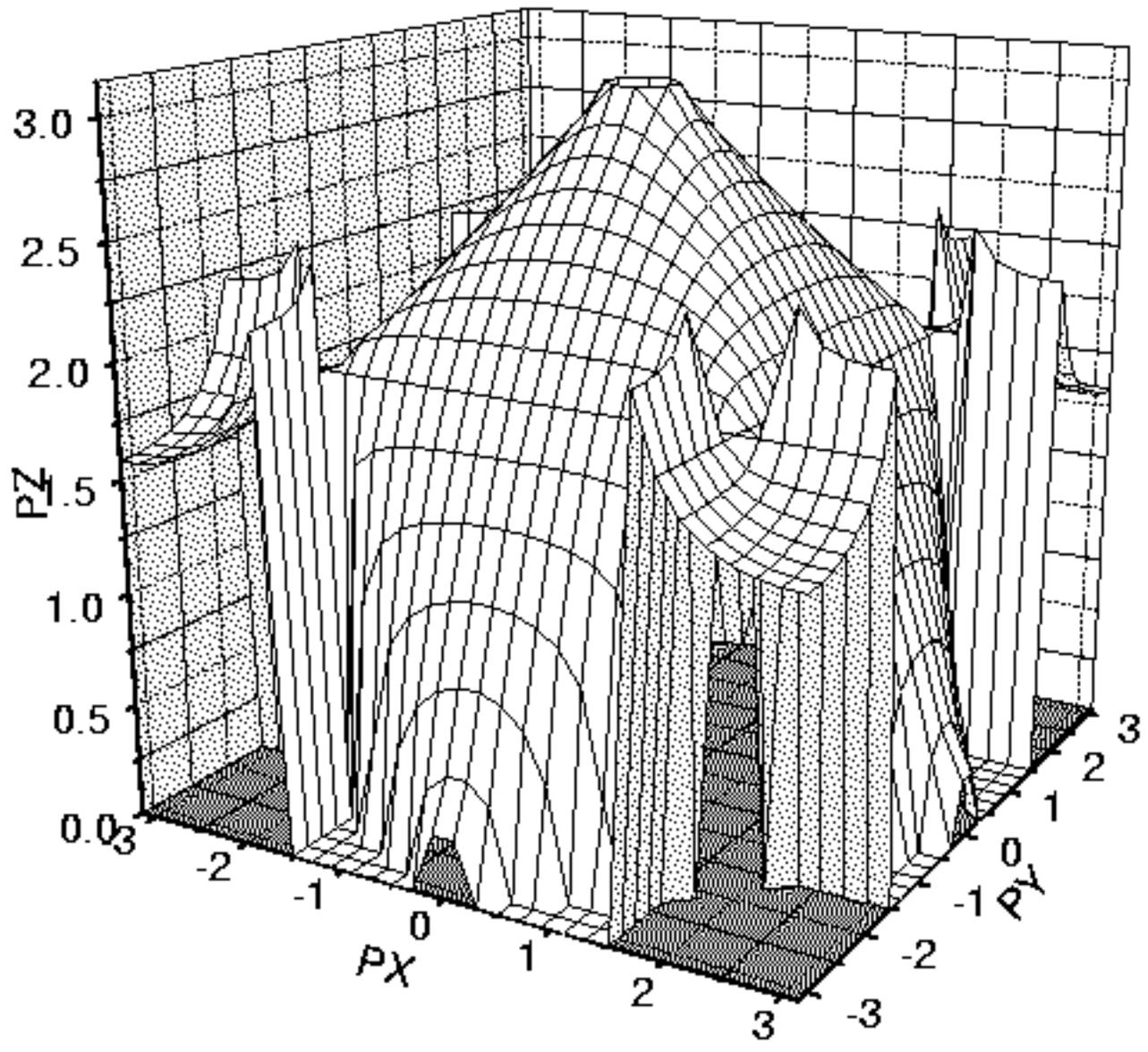
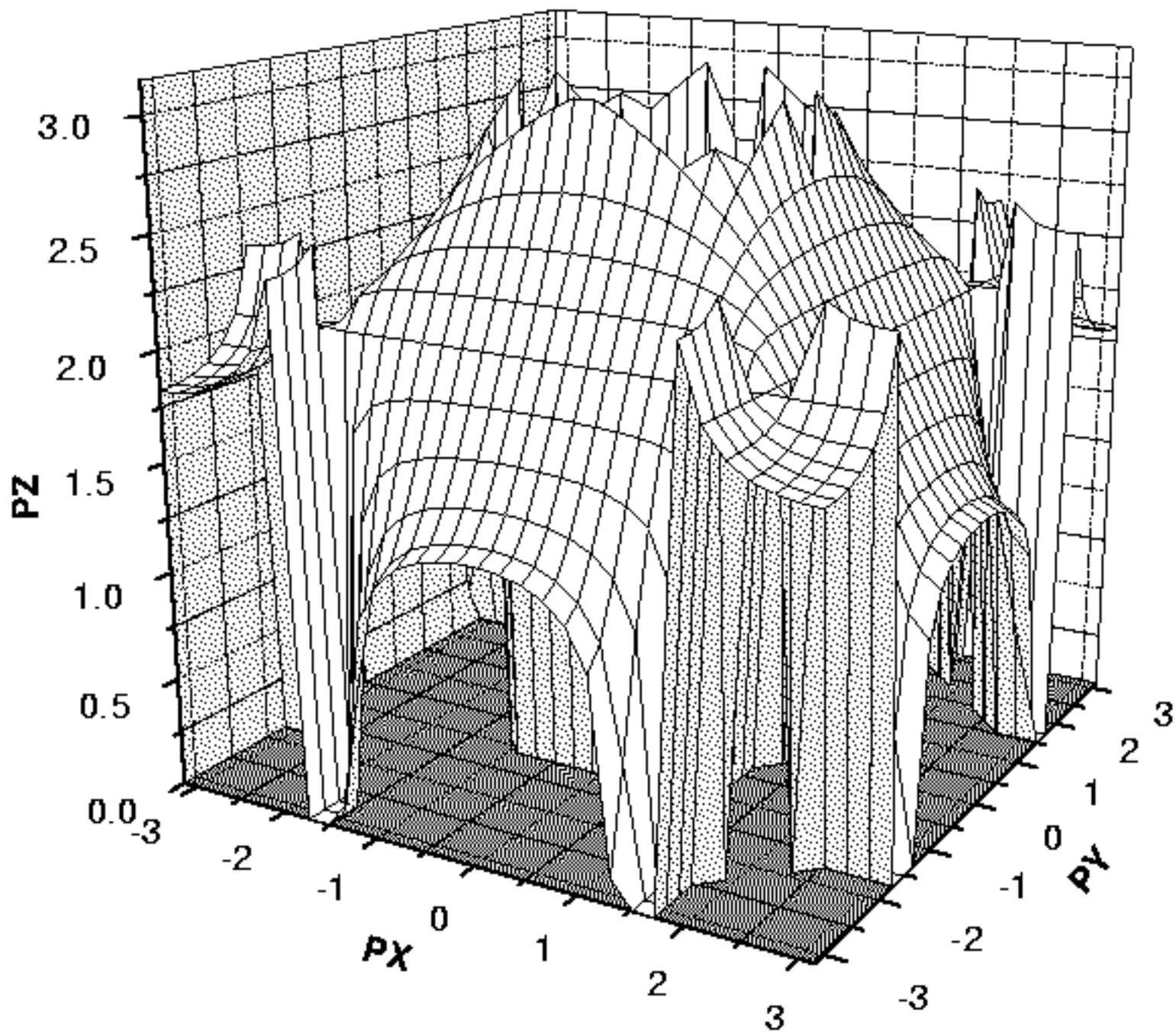
\end{document}